\documentclass[12pt]{article}
\usepackage{amsmath,amssymb}

\usepackage{graphicx}
\usepackage{wrapfig}

\usepackage{hyperref}
\usepackage{cite}

\def\beq{\begin{eqnarray}}
\def\eeq{\end{eqnarray}}

\newcommand{\Tr}{\,\mathrm{Tr}\,}            

\newcommand{\be}{\begin{equation}}
\newcommand{\ee}{\end{equation}}
\newcommand{\bea}{\begin{eqnarray}}
\newcommand{\eea}{\end{eqnarray}}
\newcommand{\bg}{\begin{gather}}
\newcommand{\eg}{\end{gather}}
\newcommand{\bseq}{\begin{subequations}}
\newcommand{\eseq}{\end{subequations}}

\renewcommand{\ln}{\mathop{\rm ln}\nolimits}

\def\tr{\hbox{Tr}}

\def\be{\begin{eqnarray}}
\def\ee{\end{eqnarray}}
\def\lb{\label}

\begin{document}

\title{\textbf{Remarks on effective action and  entanglement entropy of Maxwell field in generic gauge }}

\vspace{1cm}
\author{ \textbf{Sergey N. Solodukhin$^{\star}$  }} 

\date{}
\maketitle

\begin{center}
  \hspace{-0mm}
  \emph{Laboratoire de Math\'ematiques et Physique Th\'eorique }\\
  \emph{Universit\'e Fran\c cois-Rabelais Tours F\'ed\'eration Denis Poisson - CNRS, }\\
  \emph{Parc de Grandmont, 37200 Tours, France} \\
  \emph{and KITP, University of California, Santa Barbara, CA 93106, USA }\\    
\end{center}

{\vspace{-11cm}
\begin{flushright}
 NSF-KITP-12-167
\end{flushright}
\vspace{11cm}
}



\begin{abstract}
\noindent {   We analyze the dependence of the effective action and the entanglement entropy in  the Maxwell theory
on the gauge fixing parameter $a$ in $d$ dimensions. For a generic value of $a$ the corresponding vector operator is nonminimal. The operator can be diagonalized 
in terms of the transverse and longitudinal modes. Using this factorization we obtain an expression for the heat kernel coefficients of the nonminimal operator in terms of 
the coefficients of two minimal Beltrami-Laplace operators  acting on $0$- and $1$-forms. This expression agrees with an earlier result by Gilkey et al.  Working in a regularization scheme with the dimensionful UV regulators  we introduce three different regulators: for transverse, longitudinal and ghost modes, respectively.
We then show that the effective action and the entanglement entropy do not  depend on the gauge fixing parameter $a$ provided  
the certain ($a$-dependent)  relations are imposed on the regulators.
Comparing the entanglement entropy with the
black hole entropy expressed in terms of the induced Newton's constant we conclude that their difference, the so-called Kabat's contact term, does not depend on the gauge fixing parameter $a$. We consider this as  an indication of   gauge invariance of the contact  term.}
\end{abstract}

\vskip 2 cm
\noindent
$^{\star}$  e-mail: Sergey.Solodukhin@lmpt.univ-tours.fr

\newpage
    \tableofcontents
\pagebreak

\newpage

\section{ Introduction}

In Quantum Field Theory  there are  two well-defined  quantities  that are  sensitive to the UV behavior of the theory.
One of them is the effective action (for a nice and informative review see \cite{Barvinsky:1985an}) and the other
is the entanglement entropy (the  various approaches are recently reviewed in  \cite{ee}, 
\cite{Casini:2009sr}, \cite{Solodukhin:2011gn}, \cite{Jacobson:2012yt}). Provided a regularization scheme  with a regularization parameter $\epsilon$ is used
to handle the UV divergences the both quantities are crucially dependent on the parameter $\epsilon$ thus revealing in this dependence
the short-distance behavior of the theory. In the case of the effective action there exist some regularization schemes in which  the regularization parameter
is dimensionless. The divergences of the action when $\epsilon$ is taken to zero then correspond to the logarithmic UV divergences in those schemes, where
a dimensionful $\epsilon$ is introduced. Otherwise, with a dimensionful $\epsilon$, the effective action shows a series of the power-law UV divergent terms.
The existing tradition tends to view the power-law UV divergences as spurious and less important than the logarithmic ones, the latter are considered to be ``physical''
while the former are viewed as ``scheme dependent''.  This is especially astonishing  taking that namely a power law UV  divergent term in the effective action
produces that enormous contribution to the  cosmological constant which leads to the well known ``cosmological constant problem''.

 The situation with the entanglement entropy is quite the opposite.
Here the leading and the most important contribution to the entropy is given by the area of the entangling surface. So that a dimensionful UV regulator is natural
to be present in order to, in a combination with the area,   produce a dimensionless quantity.  Thus, in this case the power-law UV divergences  are viewed as physically important.

The different attitude towards the power law UV divergences demonstrated in these two cases is  surprising since, for all  fields minimally coupled to gravity,
there is a correspondence between the  UV divergences in the effective action and the entropy and, if we are talking about the  entanglement entropy of horizons,
the renormalization of one quantity automatically leads to the renormalization of the other \cite{Susskind:1994sm}, \cite{Fursaev:1994ea}. 
Moreover, by changing the short-distance behavior of the theory one can see that the both quantities demonstrate, in parallel, the corresponding modifications
in the structure of the UV divergences \cite{Nesterov:2010yi}.  The whole issue of the power law UV terms in the effective action becomes especially important in the models of induced gravity, where both the cosmological and Newton's constants and the black hole entropy are induced in quantum loops of some fundamental constituents \cite{ted}, \cite{Frolov:1997up}.

By taking seriously the power-law UV divergences in the effective action we, however, may have a situation, where some self-obvious properties of the effective action are less transparent. 
In particular, the usual demonstration of gauge invariance of the effective action should be re-derived\footnote{In particular, the use of Ward's identities may fail for the power-law divergent terms in the effective action \cite{Andrei}.} in this case and, possibly, supplemented by imposing some additional requirements.

In the present paper we consider these issues in the case of the Maxwell theory.  Following the standard gauge fixing procedure, we add a generic gauge fixing term to the action.
Our goal then is to demonstrate that the resultant effective action  and the entropy do not actually depend on the value of the gauge fixing parameter $a$. 
This is not an obvious property of the theory since, 
for generic values of the  gauge fixing parameter, $a\neq 1$,   the corresponding vector field operator  is nonminimal. The separation of vector modes on the transverse and longitudinal helps to
diagonalize the operator and effectively reduce the problem to the one in the minimal gauge $a=1$. That the effective action and the entanglement entropy do not depend on parameter $a$, however,
is not achieved automatically and requires us to impose some additional conditions on the regularization parameters which regularize the UV divergences in each sector: transverse, longitudinal and ghost. That one has to introduce different UV regulators for the transverse and longitudinal modes was earlier proposed
by Kabat \cite{Kabat:1995eq} in two dimensions. The relations which we derive for the UV regulators in the different sectors of the theory are consistent  with those proposed in \cite{Kabat:1995eq}. The Maxwell field can be embedded into various supersymmetric multiplets for which the entropy has been recently analyzed in \cite{Sen}.
We believe our analysis should be important for this class of theories too.

\section{Maxwell theory, gauge fixing and nonminimal \  operator}
We start with the standard action for the Maxwell theory with a generic gauge fixing term 
\be
W=\int d^x \sqrt{g}\left( \frac{1}{4}F_{\mu\nu}F^{\mu\nu} +\frac{a}{2}(\nabla_\mu A^\mu)^2\right)\, .
\lb{1}
\ee
The  corresponding field equation for the vector field $A_\mu$,
\be
\Delta^{\mu}_{(a)\ \nu}A^\nu=0\, ,
\lb{2}
\ee
is governed by a nonminimal operator
\be
\Delta^\mu_{(a)\ \nu}=- \Box_1 \delta^\mu_\nu+(1-a)\nabla^\mu\nabla_\nu+R^\mu_{\ \nu}\, .
\lb{3}
\ee
We define $\Box_k=\nabla_\alpha\nabla^\alpha$ as the Laplace operator acting on $k$-forms.
The operator (\ref{3}) becomes minimal if $a=1$. Below we will consider the general case, when
$a$ is arbitrary\footnote{The unitarity requires that $a$ to be positive. So no any other restriction 
will be imposed on $a$.} parameter. It should be note that in  the coordinate invariant form the non-minimal operator (\ref{3}) can be represented as follows
\be
\Delta_a=-(\delta_2 d_1+a\, d_0\delta_1)\, ,
\lb{3*}
\ee
where $d_k$ is the differential exterior operator acting on $k$-forms and $\delta_{k+1}$ is its adjoint operator acting on $(k+1)$-forms with the well known properties
$d_k d_{(k-1)}=0$ and $\delta_{(k+1)}\delta_k=0$. If $a=1$ then $\Delta_1=-(d_0\delta_1+\delta_2 d_1)$ is the Beltrami-Laplace operator on $1$-forms,
\be
\Delta^\mu_{1\ \nu}=- \Box_1 \delta^\mu_\nu+R^\mu_{\ \nu}\, .
\lb{BL}
\ee

The action (\ref{1}) should be supplemented by a ghost action
\be
W_{gh}=\int d^d x\sqrt{g} \frac{1}{2}\bar{c}\, \Delta_{gh}\, c\, ,
\lb{4}
\ee
where the ghost operator is
\be
 \Delta_{gh}=-\sqrt{a}\, \Box_0\, .
 \lb{5}
 \ee
So that the quantum partition function, provided the generic gauge is imposed, of the theory reads
\be
Z(a)={\det}^{-1/2}\Delta_{(a)}\, \det\Delta_{gh}\, .
\lb{6}
\ee
In our analysis of how (\ref{6}) depends on the gauge fixing  parameter $a$  it will be, however, useful to separate the contributions of the transverse and longitudinal modes in the partition function (\ref{6}).

\section{Projectors and decomposition of nonminimal operator}
Let us introduce the longitudinal and transverse mutually orthogonal projectors
\be
&&{\cal P}^\mu_{\ \nu}=\nabla^\mu\frac{1}{\Box_0}\nabla_\nu\, , \, \, \Pi^\mu_{\ \nu}=\delta^\mu_{\ \nu}-{\cal P}^\mu_{\ \nu}\, \nonumber \\
&&{\cal P}\ {\cal P}={\cal P}\, , \, \, { \Pi}\ {\Pi}={\Pi}\, , \, \, {\cal P}\ { \Pi}={ \Pi}\ {\cal P} =0\, ,
\lb{6*}
\ee
where $\Box_0$ is the covariant Laplace operator acting on scalars ($0$-form), $\Box_0=\nabla_\alpha \nabla^\alpha$. They act on vector field $A^\nu$ as $({\cal P}A)^\mu={\cal P}^\mu_{\ \nu} A^\nu$. Similarly, we can consider the Ricci tensor $R^\mu_{\ \nu}$ as operator acting on vector fields, $(RA)^\mu=R^\mu_{\ \nu}A^\nu$.
Clearly, the operator $\Pi$ is projector   onto transverse vector fields, $\nabla_\mu \, (\Pi^\mu_{\ \nu}A^\nu)=0$.

We note the useful commutation relations between the covariant Laplace operator $\Box_k$ and the covariant derivative
\be
&&\nabla^\mu\Box_0=\Box_1\nabla^\mu-R^\mu_{\ \nu}\nabla^\nu\, , \nonumber \\
&&\nabla_\mu \Box_1=\Box_0\nabla_\mu+\nabla_\alpha R^\alpha_{\ \mu}\, ,
\lb{6**}
\ee
where in the first line the operators are acting on scalars while in the second they are acting on vectors and the Ricci tensor is viewed as a matrix  operator.
Using these relations we find
\be
\nabla^\mu\nabla_\nu=\Box_1{\cal P}^\mu_{\ \nu}-R^\mu_{\ \alpha}{\cal P}^\alpha_{\ \mu}\, .
\lb{6+}
\ee
So that the non-minimal operator (\ref{3}) can be presented in the form
\be
\Delta^\mu_{(a)\ \nu}=\Delta^\mu_{1\ \alpha}\Pi^\alpha_{\ \nu}+a\, \Delta^\mu_{1\ \alpha}{\cal P}^\alpha_{\ \nu}\, ,
\lb{6&&}
\ee
where $\Delta_1$ is the minimal operator (\ref{BL}).
It is now natural  to define the transverse and longitudinal Laplace operators
\be
\Delta_T=\Delta_1\Pi\, , \, \, \Delta_L=\Delta_1{\cal P}\, 
\lb{6++}
\ee
so that we have
\be
\Delta_a=\Delta_T+a\, \Delta_L\, .
\lb{6-}
\ee
In terms of the covariant differential operators $d$ and $\delta$ the projector $\cal P$ takes the form
\be
{\cal P}=d_0\frac{1}{\Box_0}\delta_1\, ,
\lb{6--}
\ee
we remind that $\Box_0=\delta_1 d_0$. Using this representation  and the invariant form (\ref{3*}) of the operator $\Delta_a$ we find that
the operators $\Delta_T$ and $\Delta_L$ are, in fact, local,
\be
\Delta_T=-\delta_2 d_1\, , \, \, \Delta_L=- d_0\delta_1\, .
\lb{LT}
\ee
Using these invariant representations
it is easy to demonstrate the commutation relations
\be
[\Delta_1, {\cal P}]=0\, , \, \, [\Delta_T,\Delta_L]=0\, .
\lb{7-}
\ee
Moreover, since ${\cal P}$ commutes with operator $\Delta_1$ and that  $\cal P$ and $\Pi$ are orthogonal then the product of any powers of operators $\Delta_T$ and $\Delta_L$ is nul,
\be
\Delta_L^k \cdot \Delta_T^n=0\, , \, \, k,n>0\, .
\lb{kn}
\ee
These properties indicate that the determinant  of the operator $\Delta_a$, defined as product of its non-zero eigen values,  reduces to a product of two determinants 
\be
\det \Delta_a=\det \Delta_T \det (a\Delta_L)\, .
\lb{det}
\ee
This representation should be used in (\ref{6}) when we compute the partition function of the Maxwell field.

\section{Heat kernel  and small $s$ expansion}
The technical tool to be used in this paper in order to evaluate the partition function (\ref{6}) is the heat kernel. For a vector operator ${\cal D}^\mu_{\ \nu}$ the heat kernel  $K^\mu_{\ \nu}(s,x,x')$   satisfies the equation
\be
\partial_s K^\mu_{\ \nu}(s, x,x')+{\cal D}_{\ \sigma}^\mu(x)\, K^\sigma_{\ \nu}(s,x,x')=0
\lb{7}
\ee
and the ``initial''  condition
\be
K^\mu_{\ \nu}(s=0,x,x')=\delta^\mu_\nu\, \delta^{(d)}(x,x')\, .
\lb{8}
\ee
With the help of the heat kernel we can express the effective action $W_{eff}=-\frac{1}{2}\ln \det {\cal D}$ as follows
\be
W_{eff}=-\frac{1}{2}\int^\infty_{\epsilon^2}\frac{ds}{s}\int d^d x\, \Tr K(s, x'=x)\, ,
\lb{9}
\ee
where $\epsilon$ is an UV cut-off. The heat kernel is characterized by its small $s$ expansion,
\be
\Tr K(s,x=x')=\frac{1}{(4\pi )^{d/2}}\sum_{k=0} c_k({\cal D})s^{\frac{2k-d}{2}}\, .
\lb{9'}
\ee

A formal solution to equation (\ref{7}) is $K=e^{-s{\cal D}}$. For the nonminimal operator $\Delta_a$ the heat kernel reduces, as can be seen by using properties (\ref{7-}) and (\ref{kn}), to a sum of two heat kernels
\be
e^{-s\Delta_a}&=&e^{-s\Delta_T}+e^{-sa\Delta_L}\nonumber \\
&=&e^{-s\Delta_1}\Pi+e^{-sa\Delta_1}{\cal P}\, .
\lb{heat}
\ee
This relation can be obtained by first using the representation of exponential $e^{-s\Delta_a}$ as Taylor series and then using the commutation relations (\ref{7-}) and the orthogonality property (\ref{kn}). 
Consider now a small $s$ expansion of the heat kernels in  (\ref{heat}). We find a relation
\be
c_k(\Delta_a)=c_k(\Delta_T)+a^{\frac{2k-d}{2}}c_k(\Delta_L)\, .
\lb{ck}
\ee
Operator $\Delta_L=-d_0\delta_1$ has same non-zero eigen values as the scalar Laplace operator $-\Box_0=-\delta_1d_0$, hence one has that $c_k(\Delta_L)=c_k(-\Box_0)$.
On the other hand, if $a=1$ then $c_k(\Delta_1)=c_k(\Delta_T)+c_k(\Delta_L)$ from which we find that $c_k(\Delta_T)=c_k(\Delta_1)-c_k(-\Box_0)$.
Putting everything together  we express the heat kernel coefficients of the non-minimal operator $\Delta_a$ in terms of the coefficients of the heat kernel of two minimal operators,
\be
c_k(\Delta_a)=c_k(\Delta_1)+(a^{\frac{2k-d}{2}}-1)c_k(-\Box_0)\, .
\lb{ckk}
\ee
This relation agrees with an earlier result obtained  in \cite{Gilkey}. We remind the reader the first few heat kernel coefficients of the minimal operators,
\be
&&c_0(-\Box_0)=\sqrt{g}\, ,\, \, c_1(-\Box_0)=\sqrt{g}\, \frac{1}{6}R\, ,\nonumber \\
&&c_0(\Delta_1)=\sqrt{g}\, d\, , \, \, c_1(\Delta_1)=\sqrt{g}\, (\frac{d}{6}R-R)\, .
\lb{cc}
\ee
We have independently checked the relation (\ref{ckk}) for coefficients with $k=0,1$ by using the momentum space method developed in \cite{Nesterov:2010yi}
(details of this calculation are  not included here and are available upon request) and we have  obtained a complete agreement with (\ref{ckk}).

\section{A formal demonstration  of gauge independence}

Let us start with a formal demonstration that the partition function  (\ref{6}) does not depend on the gauge fixing parameter $a$. Our starting point is the variation formula for the determinant. This formula can be obtained by varying the identity $\ln \det {\cal D}=\Tr \ln {\cal D}$, valid for any operator $\cal D$,
\be
\delta \ln \det {\cal D}=\Tr {\cal D}^{-1}\delta {\cal D}\, ,
\lb{f1}
\ee
where ${\cal D}^{-1}$ is the inverse operator. 
It can be represented by integral over proper time
\be
{\cal D}^{-1}=\int_{\epsilon^2}^\infty dse^{-s{\cal D}}\, ,
\lb{f2}
\ee
where we introduced a regulator $\epsilon$.  In what follows we introduce  a separate regulator for each operator in question: $\epsilon_T$, $\epsilon_L$ and $\epsilon_G$.

For the operators at hand we have the following variation with respect to parameter $a$,
\be
\delta \Delta_a=\delta a \, (-d_0\delta_1)\, , \, \, \, \delta \Delta_{gh}=\frac{\delta a}{2\sqrt{a}}(-\Box_0)\, .
\lb{f3}
\ee
For the ghost operator we find 
\be
\Tr\Delta^{-1}_{gh}\delta \Delta_{gh}=-\frac{\delta a}{2\sqrt{a}}\int_{\epsilon^2_{G}}^\infty ds \Tr( e^{-s\Delta_{gh}}\Box_0)=-\frac{\delta a}{2a}\int_{\epsilon^2_{G}\sqrt{a}}^\infty ds \Tr( e^{s\Box_0}\Box_0)\, .
\lb{f4}
\ee 
Then, using that $e^{-s\Delta_a}d_0\delta_1=(e^{s\delta_2 d_1}+e^{asd_0\delta_1})d_0\delta_1=e^{asd_0\delta_1}d_0\delta_1$, we obtain for the vector non-minimal operator that
\be
\Tr\Delta^{-1}_a\delta \Delta_a=-\delta a\int_{\epsilon_L^2}^\infty ds \Tr(e^{as d_0\delta_1} d_0\delta_1)=-\frac{\delta a}{a}\int^\infty_{\epsilon^2_La}ds \Tr (e^{s\Box_0}\Box_0)\, ,
\lb{f5}
\ee
where we introduced a separate regulator for the longitudinal operator, rescaled the proper time and, finally,  used the cyclicity property of the trace, $\Tr (d_0\delta_1)^n=\Tr (\delta_1 d_0)^n$  and
\be
\Tr (e^{sd_0\delta_1}d_0\delta_1)=\Tr (e^{s\delta_1 d_0}\delta_0 d_1) =\Tr (e^{s\Box_0}\Box_0)\, .
\lb{f6}
\ee 

In a even more formal analysis one could replace  the lower limit in the integrals (\ref{f4}), (\ref{f5}) by zero. Then one would immediately conclude that the two variations (\ref{f4}) and (\ref{f5})
mutually cancel each other in the variation of the partition function (\ref{6}). This is that sort of reasoning which is usually used to demonstrate  the gauge independence of the partition function.

In the presence of the UV regulators the cancellation of two contributions in the variation of the partition function (\ref{6}) with respect to $a$   is still possible if
we impose a certain relation between two regulators, 
\be
\epsilon^2_G=\epsilon_L^2\sqrt{a}\, .
\lb{f7}
\ee

This, still rather formal, demonstration can not be correct if the power law UV divergences are present, as we discuss below. A possible reason for this, mentioned to us by A. Barvinsky \cite{Andrei}, is the failure of the cyclicity property of trace, used in (\ref{f6}),
in the presence of the power law UV divergent terms.

\section{Effective action and UV regulators, gauge independence}
Let us now return to the heat kernel representation of the  total effective action of the Maxwell theory and separate the three different contributions: transverse, longitudinal and ghost, 
\be
W_{Maxwell}=-\frac{1}{2}\left(\int^\infty_{\epsilon_T^2}\frac{ds}{s}\Tr e^{-s\Delta_T}+\int^\infty_{\epsilon_L^2}\frac{ds}{s}\Tr e^{-sa\Delta_L}-2\int^\infty_{\epsilon_G^2}\frac{ds}{s}\Tr e^{-s\Delta_{gh}}\right)\, .
\lb{Maxwell}
\ee
We remind that the ghost operator $\Delta_{gh}=-\sqrt{a}\Box_0$.

The relation (\ref{f7}) guarantees that dependence on $a$ is not present in the 
logarithmic terms, proportional to $\ln\epsilon_L$ and $\ln\epsilon_G$, and in the UV finite terms. If, however,   there is a power law  divergence of the type  $1/\epsilon^n$
in $\ln \det$ of operators $\Delta_{gh}$ and $\Delta_L$ then the relation (\ref{f7}) is not enough to guarantee that a power law dependence on $a$ in these terms is absent.
This is due to the fact that these terms come with a relative  coefficient $-1/2$  in (\ref{Maxwell}). 
This relative coefficient is important for the cancellation in the logarithmic terms but does not help at all in the case of  the power law divergent terms. 
Thus, the idea, which is behind the formal analysis in the previous section,  that the dependence on the parameter $a$  may disappear due to the mutual cancellation of contributions of the longitudinal and the ghost parts does not work.  The only possibility then is that each of the three terms in (\ref{Maxwell}), individually, should  not depend on $a$.

The important point here is that we introduced three different UV regulators, $\epsilon_T$, $\epsilon_L$ and $\epsilon_G$, for each operator involved.
They are not, however, arbitrary. Suppose we choose $\epsilon_T=\epsilon$ as our benchmark. This parameter regularizes the operator $\Delta_T$ or, if $a=1$, the operator $\Delta_1$.
In flat space $\Delta_1=-\partial^2$ is just a diagonal product of two derivatives. Then, in order to regularize an  operator $-a^k\partial^2$ we have to use the regulator $\epsilon/a^{k/2}$.
Thus, we conclude that we have to choose
\be
\epsilon_T=\epsilon\, , \, \, \epsilon^2_L=\epsilon^2/a\, , \, \, \epsilon^2_G=\epsilon^2/\sqrt{a}\, .
\lb{epsilon}
\ee
The relation between $\epsilon_T$ and $\epsilon_L$ is consistent with the analysis made by   Kabat \cite{Kabat:1995eq} in two dimensions on the basis of the BRST invariance.

We note that (\ref{epsilon}) includes the relation (\ref{f7}) which we derived   on the basis of the formal arguments. The opposite however is not true.
That is why (\ref{f7}) alone is not enough to remove all dependence on $a$ while this can be done using (\ref{epsilon}). 
With the choice (\ref{epsilon}) for the regulators the effective action (\ref{Maxwell}) becomes completely independent of the gauge fixing parameter $a$ and, moreover,  it is the same as in the minimal gauge $a=1$,
\be
W_{Maxwell}=-\frac{1}{2}\int^\infty_{\epsilon^2}\frac{ds}{s}(\Tr e^{-s\Delta_1}-2\tr e^{+s\Box_0})\, .
\lb{Maxwell1}
\ee
We note that  if there exist some zero modes of the operators in question the discussion should be supplemented by subtraction of the contribution of these zero modes (see \cite{Barvinsky:1995dp} and  \cite{Donnelly:2012st} for a discussion in two dimensions). This subtraction modifies the logarithmic UV divergent 
terms in the action.

\section{Cosmological and Newton's constants}

Using the heat kernel  coefficient (\ref{cc})  one calculates the power law divergences in the effective action (\ref{9}) on the curved background,
\be
&&W_{eff}=-\lambda(\epsilon)\int d^d x \sqrt{g}-\frac{1}{16\pi G(\epsilon)}\int d^d x\sqrt{g}R\, ,
\lb{W1}
\ee
where
\be
\lambda(\epsilon)=-\frac{1}{d(4\pi)^{d/2}}\frac{1}{\epsilon^{d}}(d-2)\,
\lb{L1}
\ee
is the induced cosmological constant and
\be
\frac{1}{16\pi G(\epsilon)}=\frac{1}{(d-2)}\frac{1}{(4\pi)^{d/2}}\frac{1}{\epsilon^{d-2}}\left(\frac{1}{6}(d-2)-1\right)\, 
\lb{G1}
\ee
is the induced Newton's constant. We notice that $(d-2)=N_{s=1}(d)$ is the number of on-shell degrees of freedom of a spin $s=1$ particle  in dimension $d$.

\section{Minkowski spacetime: heat kernel in momentum space}
In Minkowski spacetime one can use the representation of the heat kernel by means of the Fourier transform. For the heat kernel of the nonminimal operator (\ref{3})
one has that
\be
K^\mu_{\ \nu}(s,x,x')=\frac{1}{(2\pi)^{d}}\int d^p K^\mu_{ \ \nu}(s,p)e^{-ip(x-x')}\, ,
\lb{10}
\ee
where $K^\mu_{ \ \nu}(s,p)$ satisfies equation
\be
\partial_s\, K^\mu_{\ \nu}(s,p)+(p^2\delta^\mu_\sigma+(a-1)p^\mu p_\sigma)K^\sigma_{\ \nu}(s,p)=0\, 
\lb{11}
\ee
and the initial condition
$
K^\mu_{\ \nu}(s=0,p)=\delta^\mu_\nu\, .
$
The  solution is easily found,
\be
K^\mu_{\ \nu}(s,p)=(e^{-s\Delta_a})^\mu_{\ \nu}=(\delta^\mu_\nu-\frac{p^\mu p_\nu}{p^2})e^{-p^2s}+\frac{p^\mu p_\nu}{p^2}e^{-ap^2s}\, .
\lb{13}
\ee
This is exactly the form (\ref{heat}), where in the momentum space the projectors are defined as 
\be
{\cal P}^\mu_{\ \nu}=\frac{p^\mu p_\nu}{p^2}\, , \, \, {\Pi}^\mu_{\ \nu}=(\delta^\mu_\nu-\frac{p^\mu p_\nu}{p^2})\, .
\lb{14}
\ee
Correspondingly, for the heat kernel of the ghost operator we have that
\be
K^G(s,p)=e^{-s\sqrt{a}p^2}\, .
\lb{15}
\ee

\section{Entanglement entropy}\lb{entropy}

On a hypersurface of constant time in Minkowski spacetime we consider a co-dimension two surface $\Sigma$. Entanglement entropy is defined by tracing over the modes which reside inside the surface. Provided one starts with a pure quantum state after the tracing over one ends up with a mixed state characterized by  a density matrix. The corresponding entropy is called 
entanglement entropy. In oder to calculate the entanglement entropy one usually uses the so-called replica trick. It consists in introducing a small angle deficit 
$\delta=2\pi(1-\alpha)$ at the surface and differentiating the effective action computed on this conical space with respect to the angle deficit (see \cite{Solodukhin:2011gn} for a review). 
This procedure may be complicated in general. In some simple cases, when, for example, the surface $\Sigma$ is infinite $(d-2)$-plane, the procedure is rather straightforward.
First, we need to compute the trace of the heat kernel on the conical space. In the momentum space representation of the heat kernel this procedure was carried out in general
in \cite{Nesterov:2010yi}. The result can be formulated as follows
\be
\Tr K_\alpha(s)=\frac{1}{(4\pi)^{d/2}}\left(\alpha V P_{d}(s)+\frac{\pi}{3\alpha}(1-\alpha^2) A(\Sigma)P_{d-2}(s)\right)\, ,
\lb{24}
\ee
where we introduced 
\be
P_n(s)=\frac{2}{\Gamma(\frac{n}{2})}\int_0^\infty dp p^{n-1}\Tr K(s,p)\, ,
\lb{25}
\ee
$K(s,p)$ is the Fourier transform of the heat kernel in question.
The normalization is chosen in such a way that for $\Tr K(s,p)=e^{-sp^2}$ we have that $P_n(s)=s^{-{n/2}}$.
$A(\Sigma)$ is the area of the surface $\Sigma$. The second term in (\ref{24}) is proportional to the angle deficit and, thus, contributes to the entropy.
Entanglement entropy then  takes the form \cite{Nesterov:2010yi}
\be
S=\frac{A(\Sigma)}{12 (4\pi)^{(d-2)/2}}\int_{\epsilon^2}^\infty\frac{ds}{s}P_{d-2}(s)\, .
\lb{26}
\ee

Applying this general formula to the Maxwell theory with  a generic gauge fixing term we have to consider separately  the contributions of transverse, longitudinal and ghost modes:
\be
&&P^{(T,L,G)}_{d-2}(s)=\frac{2}{\Gamma(\frac{d-2}{2})}\int_0^\infty dp p^{d-3}\Tr K^{(T,L,G)}(s,p)\, , \nonumber \\
&&\Tr K^T(s,p)=\Tr e^{-sp^2}\Pi\, , \, \, \Tr K^L(s,p)=\Tr e^{-asp^2}{\cal P} \, , \, \, \Tr K^G(s,p)=e^{-\sqrt{a}sp^2}\, .
\lb{KK}
\ee
Since $\Tr \Pi=(d-1)$ and $\Tr {\cal P}=1$   we find
\be
P^T_{d-2}(s)=(d-1)s^{-(d-2)/2}\, , \, \, P^L_{d-2}(s)=(as)^{-(d-2)/2}\, , \, \, P^G_{d-2}(s)=(\sqrt{a}s)^{-(d-2)/2}\, .
\lb{PP}
\ee
The entanglement entropy is sum of the contributions of each sector of the Maxwell theory,
\be
S=\frac{A(\Sigma)}{12(4\pi)^{(d-2)/2}}\left(\int_{\epsilon_T^2}^\infty\frac{ds}{s}P^T_{d-2}(s)+\int_{\epsilon_L^2}^\infty\frac{ds}{s}P^L_{d-2}(s)-2\int_{\epsilon_G^2}^\infty\frac{ds}{s}P^G_{d-2}(s)\right)\, ,
\lb{SS}
\ee
where the ghost modes give a negative contribution and each of three sectors (transverse, longitudinal and ghost) has its own regularization parameter as in the case of the effective action.
For the choice (\ref{epsilon}) of the regulators the entanglement entropy clearly does not depend on the gauge fixing parameter $a$ and it equals
\be
S_{ent}=\frac{A(\Sigma)}{6(d-2)(4\pi)^{\frac{d-2}{2}}\epsilon^{d-2}}(d-2)\, .
\lb{29}
\ee
As before, we notice that $(d-2)=N_{s=1}(d)$ is the number of the on-shell degrees of freedom of a spin $s=1$ particle in $d$ dimensions. 

We conclude that the prescription (\ref{epsilon}), which makes the effective action independent of the gauge fixing parameter $a$, automatically leads to the entanglement entropy which does not depend on $a$. This adds yet another aspect to the correspondence between the effective action and the entropy earlier studied in \cite{Fursaev:1994ea}.  
Our analysis in this section has been made in Minkowski spacetime. However, the  result for the entanglement entropy  in flat spacetime is always the same as the leading UV divergent contribution (proportional to the area)
to the entanglement entropy in curved spacetime. Thus, the entropy (\ref{29}) shows the area law for the Maxwell theory in a curved  spacetime.

\section{Kabat's contact term}
Black holes in a theory of gravity described by the gravitational action (\ref{W1}) have the Bekenstein-Hawking entropy proportional to the horizon area
\be
S_{BH}=\frac{A}{4\pi G(\epsilon)}\, ,
\lb{BH}
\ee
measured in the units set  by  $G(\epsilon)$, the induced Newton's constant (\ref{G1}).  This should be compared to the 
entanglement entropy of the horizon that we calculated in section \ref{entropy}. Here we note that for all known matter fields, bosonic or fermionic, 
minimally coupled to gravity these two entropies are identical \cite{Fursaev:1994ea}, \cite{Solodukhin:2011gn}. The difference, however, takes place for those  fields that nonminimally couple to the background metric, i.e. couple directly to the curvature. In particular, this is so for  the gauge fields as can be seen from the form of the field equation (\ref{3}).
The difference between the two entropies can be presented as follows
\be
S_{BH}=S_{ent}-Q\, ,
\lb{2ent}
\ee
where
\be
Q=\frac{A(\Sigma)}{(d-2)(4\pi)^{\frac{d-2}{2}}\epsilon^{d-2}}\,  
\lb{Q}
\ee
is  what in the literature known as Kabat's contact term (the interest to this  term has been recently renewed, see  \cite{Zhitnitsky:2011tr}, \cite{Donnelly:2012st}, \cite{Kabat:2012ns}). It appears due to a direct interaction of the gauge field with the tip of the cone.
For a nonminimally coupled scalar field, where a similar phenomenon takes place, this was analyzed in \cite{Solodukhin:1995ak}.
The Ricci tensor on a conical space has a delta-like contribution from the conical singularity \cite{Fursaev:1995ef}
\be
R_{\mu\nu}=2\pi (1-\alpha) (n_\mu n_\nu)\delta_\Sigma +R^{reg}_{\mu\nu}\, ,
\lb{Ricci}
\ee
where $n^a_\mu$, $a=1,2$ is a pair of vectors orthogonal to $\Sigma$, $(n_\mu n_\nu)=\sum_a n^a_\mu n^a_\nu$.  The black hole entropy
(\ref{BH}) is obtained as response of the gravitational action to a small angle deficit at the horizon, so that $Q$ arises as such a response 
which is due to the coupling of the vector field directly to the Ricci tensor,
\be
Q=2\pi <A^\mu A^\nu>(n_\mu n_\nu)|_\Sigma\, .
\lb{QQ}
\ee
In the context of the induced gravity a similar quantity, due to scalar fields, was introduced in \cite{Frolov:1997up}.
This quantity does not have a statistical meaning and it just restores the balance between the entanglement entropy which has this meaning and the black hole (or conical) entropy \cite{Solodukhin:2011gn}.
Our main observation here is the following. Both the black hole entropy (\ref{BH}) and the entanglement entropy (\ref{29}), as we have just shown, do not depend on the gauge fixing parameter $a$. Same is true for their difference 
$Q$ (\ref{Q}), (\ref{QQ}).  Thus, the contact term $Q$  appears to be gauge independent. The gauge independence  of $Q$ is not  obvious if we use the representation (\ref{QQ}).
The other piece of evidence that the contact terms produce the gauge invariant effects is the observation made in  \cite{Kabat:2012ns}  that they make  observable and gauge
invariant contribution to the force between two cosmic strings.

The quantity $Q$ (\ref{QQ}) and the entanglement entropy (\ref{29}) can be defined for any co-dimension two surface $\Sigma$.
However,  the black hole entropy (\ref{BH}) is defined only
if $\Sigma$ is a horizon. Thus, only in the case,  when the entangling surface is a horizon,  our analysis indicates that the contact term $Q$ is a gauge-independent quantity.

\section{Conclusions}
As our final remark we note that the property that the effective action and the entanglement  entropy do not depend on the gauge fixing parameter $a$ is not completely equivalent to the gauge invariance of the effective action or the entropy. The two, however, are based on the use of  Ward's identities. That is why, although more work is likely needed to demonstrate, in full,  the gauge invariance of the contact term $Q$, the fact that $Q$ does not depend on the gauge fixing parameter is a clear argument in favor of the gauge invariance of $Q$.  The other evidence, as we already said, comes from the analysis made in \cite{Kabat:2012ns}. The gauge invariance of $Q$, if definitely established, should shed more light on the nature of this term and on its relevance to the problem of the black hole entropy.

\bigskip
\section*{Acknowledgements}
I thank  W. Donnelly, T. Jacobson, D. Kabat,  A. Wall and especially A. Barvinsky for discussions and inspiring  remarks. It is my pleasure to thank the Kavli Institute  for Theoretical Physics (KITP)
at UC Santa Barbara for hospitality during the initial stage of this project. This research was  supported in part by the National Science Foundation under Grant No. PHY11-25915.
The email exchange and the useful remarks by C. Pope, E. Sezgin, M. Perry and R. Percacci are highly appreciated.

\end{document}